# Diethyl Sulfoxide as a Novel Neutral Ligand in the Platinum Complex Anion

Vitaly V. Chaban

P.E.S., Vasilievsky Island, Saint Petersburg, Russian Federation.

**Abstract**. Diethyl sulfoxide (DESO) is far less known than its shorter-alkyl-chain homolog, dimethyl sulfoxide. Although the field of senior dialkyl sulfoxides does not currently exhibit an explosive growth, new fundamental and applied research works routinely appear from a few research groups in the world. Recently, the tetraethylammonium diethylsulfoxidopentachloroplatinate complex compound was synthesized containing the DESO molecule as a neutral ligand. In the present paper, we use a systematic computational method to rationalize the mentioned synthetic achievement in coordination chemistry. We show that only up to two DESO molecules may exist in the platinum (IV) complex ion, whereas all higher contents of DESO are thermodynamically unstable and the sterical factor plays an important role in their instabilities. Structural analysis of the tetraethylammonium diethylsulfoxidopentachloroplatinate ion pair reveals its rather strong cation-anion coordination and for the first time explains an experimentally derived high melting point. The reported results are expected to inspire experimental efforts to extend the universe of senior sulfoxides as neutral organic ligands in d-metal complexes.

**Graphical Abstract**

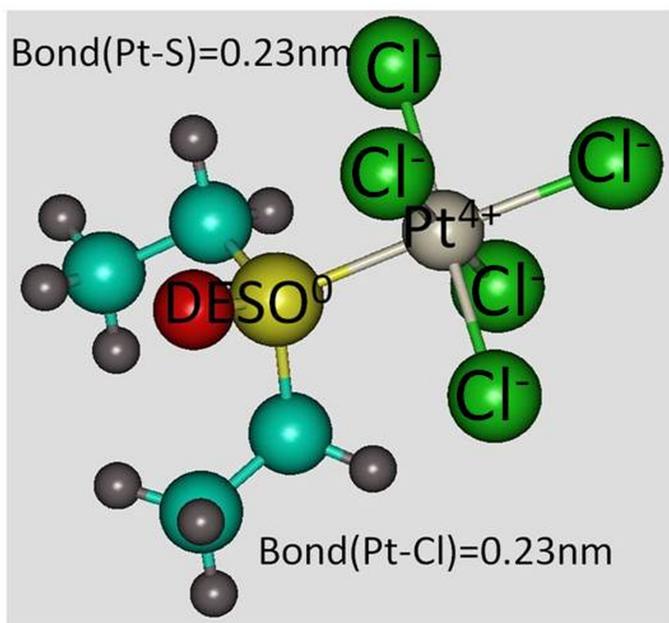

**Introduction**

Sulfoxides represent an interesting group of organic compounds that play an essential and gradually increasing role in modern chemistry.[1–8] The major homolog of the sulfoxides, dimethyl sulfoxide, has established applications in organic synthesis as a mild oxidant, in medicine as a solvent of drugs, and in biology as a rather successful cryoprotectant.[9] Senior dialkyl sulfoxides have been studied to a significantly lesser extent, although certain interesting features of their physical-chemical behavior were outlined. Markarian, Ghazoyan and coworkers reported a number of studies on the solvation properties of DESO with different co-solvents also considered it as a potential amphiphilic solvent and flexible cosolvent.[10–12] DESO-water mixtures[13–16] represent interest in view of the nascent biological applications. Skaf and coworkers[5] and Lyubartsev and coworkers[7] were among the very first researchers who made significant impacts on the reliable atomistic description using in-silico simulations of the liquid sulfoxide systems and greatly simplified their understanding at an atomistic level of precision. Furthermore, DESO was shown to have a clear potential in the cryoprotection of living cells that is seemingly even higher than that of dimethyl sulfoxide.[17]

Recently, an interesting new field of the DESO research was revealed. Tkacheva and coworkers[18] obtained a platinum complex anion and characterized its structure. The mixture of acetonitrile and DESO was applied in the synthesis of the DESO-ligand containing platinum complex.[18] Novel compounds were obtained, namely triphenylethylphosphonium diethylsulfoxidopentachloroplatinate $[Ph_3PC_2H_5][(PtCl_5(deso-S)]$ and tetraethylammonium diethylsulfoxidopentachloroplatinate $[(C_2H_5)_4N][PtCl_5(deso-S)]$. The donor-acceptor coordination occurs thanks to the presence of the sulfur atom in DESO. These orange-colored products both of which were obtained with good yields represent a fundamental importance and foster further synthetic efforts in coordination chemistry.

Just a single DESO molecule was reported to substitute one of the chlorine atoms upon the reaction of tetraorganylphosphonium and tetraethylammonium chlorides with hexachloroplatinic acid hydrate. It is, however, of natural chemical interest to find out whether a larger number of the DESO ligands can join the complex. If the chlorine atom gets replaced by the DESO molecule in the doubly negatively charged hexachloroplatinate ion, its electrostatic charge decreases proportionally, therefore providing $[PtCl_5DESO]^-$. Further exchange of ligands would provide a neutral particle $[PtCl_4DESO_2]^0$, whereas tridiethylsulfoxidotrichloroplatinate $[PtCl_3DESO_3]^+$ is expected to be a cation. Thus, a few novel and interesting ions may emerge if the discussed ligand-exchange reaction can be extended beyond a single substitution. Complex anion can be straightforwardly incorporated in versatile ionic compounds including ionic liquids and, possibly, room-temperature ionic liquids.

Platinum has a rich chemistry that includes numerous inorganic, organic, and coordination compounds, is extensively used as an efficient catalyst.[19] Moreover, platinum complexes are actively used in medicine. For instance, at least half of the patients with cancer who receive chemotherapy deal with the platinum complexes. The first-in-history anticancer platinum drug was cisplatin. It exerts cytotoxic effects when attaching to the DNA molecule at the adenine and guanine residues better than other anticancer drugs.[20] After successful application of cisplatin in the clinical practices, carboplatin and oxaliplatin were introduced. These remedies inhibit transcription and replication processes of the malignant cells leading them to the accelerated death.[21] Cisplatin derivatives are also actively explored clinically for the treatment of certain solid tumors and, possibly, the decrease of undesirable toxicity.[22] Although the side effects of the platinum-based anticancer are manifold, these compounds are considered very important in the treatment of tumors in the modern medicine.[20] Introduction of organic neutral ligands to the platinum complexes may lead to some interesting properties and definitely deserves a careful investigation from a chemical perspective.

In the present work, we used electronic-structure calculations, molecular dynamics simulations, the global-minimum-search procedure, and the reaction coordinate computations to investigate a possibility of adding more than one DESO ligand to the platinum complex. Furthermore, we investigated the structure of the tetraethylammonium diethylsulfoxidopentachloroplatinate system and systematically discussed its peculiarities with an atomistic precision. To convincingly link theoretical and experimental research, a point-by-point comparison of the structural and spectral results was hereby reported.

**Methodology**

Calculation of the vibrational frequencies of the optimized molecular wave function provides a spectrum in the mid- and far-infrared diapason of the wave numbers. We obtained a ground-state wave function of the system using the hybrid DFT Hamiltonian B3LYP.[23,24] The electronic orbitals of the simulated atoms were represented by two basis sets. Oxygen, hydrogen, carbon, sulfur, and chlorine atoms were represented by the gaussian functions of the polarized atom-centered split-valence triple-zeta basis set 6-311++G**.[25] Platinum was represented by the Los-Alamos-National-Laboratory-2-double-zeta basis set.[26] The system's wave function was considered converged when the energy difference between two consequent iterations were inferior to $10^{-4}$ kJ mol$^{-1}$. The molecular geometries were optimized with a help of the steepest-descent algorithm. The calculation was considered finished when all forces in the system were smaller than $5 \times 10^{-4}$ kJ mol$^{-1}$ nm$^{-1}$. Vibrational frequencies were computed for the global-minimum configuration that contained no imaginary frequencies in its profile.

Table 1 summarizes all chemical compositions that were investigated in this work. First of all, we are interested in the maximum number of the DESO molecules that can exist in the platinum complex cation, anion, or molecule. Second, we characterize the local structure in the ion pair tetraethylammonium diethylsulfoxidopentachloroplatinate and compare it to the experimental data reported by Tkacheva and coworkers.[18] Third, we provide comparison between computational and experimental data to derive quantitative and qualitative correlations where possible.

Table 1. The list of ionic and molecular systems that were simulated in the present work.

| # | n (DESO) | n (chlorine atoms) | n (cations) | Charge, e | Sampling time, ps | n (configurations) |
|---|---|---|---|---|---|---|
| 1 | 1 | 5 | 0 | -1 | 50 | 100 |
| 2 | 2 | 4 | 0 | 0 | 50 | 100 + 100 |
| 3 | 3 | 3 | 0 | +1 | 150 | 100 + 150 |
| 4 | 3 | 4 | 0 | 0 | 150 | 100 + 150 |
| 5 | 4 | 4 | 0 | 0 | 25 | 50 |
| 6 | 5 | 6 | 0 | -2 | 25 | 50 |
| 7 | 6 | 6 | 0 | -2 | 25 | 50 |
| 8 | 6 | 4 | 0 | 0 | 25 | 50 |
| 9 | 6 | 4 | 0 | 0 | 25 | 50 |
| 10 | 1 | 5 | 1 | 0 | 200 | 100 |
| 11 | 2 | 10 | 2 | 0 | 200 | — |
| 12 | 4 | 8 | 0 | 0 | 200 | 100 |

Stable (stationary point) configurations of systems presented in Table 1 were obtained as described herein. The systems were equilibrated using PM7-MD simulations[27–33] equipped with the Andersen temperature bath coupling (reference temperature of 300 K and collision frequency of 0.05).[34] The integration time-step in the Verlet routine was set to $5\times10^{-4}$ ps. The equilibration process was controlled by the block-averaged data of kinetic energy, potential energy, total energy, and temperature. Upon the production run, the kinetic energy equivalent to 1000 K was injected into the system every 0.25-1.00 ps. This energy was assigned randomly to atoms according to the Maxwell-Boltzmann distribution in addition to their genuine velocities. Additional portions of kinetic energy is needed for the system to overcome possible potential barriers and sample all possible states of the phase trajectory. After the system accommodated an external portion of energy, an immediate atom configuration was recorded. Subsequently, the selected configurations were optimized using the PM7 Hamiltonian[27,35,36] to remove significant interatomic forces. An ultimate geometry optimization was performed at the B3LYP/6-31G* level of theory by means of the steepest-descent method whose threshold was set to $1\times10^{-3}$ kJ mol$^{-1}$ nm$^{-1}$ (maximum force) and 0.1 kJ mol$^{-1}$ (potential energy). The empirical dispersion correction[37] to get more accurate non-covalent interactions and compensate for a possibly excessive electron delocalization in the system was performed.

Relaxed energy scans were performed through the stepwise imposing of a single distance constraint between the two selected non-bonded atoms, whereas all other atoms were free to displace in response to the acting forces (gradients). The restrained geometry optimizations upon the scan were performed according to the steepest-descent algorithm with a convergence criterion of 0.1 kJ mol$^{-1}$.

Radial distribution functions (RDFs) were obtained from conventional PM7-MD trajectories at 300 K (Andersen thermostat) with the integration time-step of equations of motion equal to $5\times10^{-4}$ ps. The trajectory frames were saved every 0.05 ps for further analysis with the in-home utilities.

Overall, more than a million self-consistent field calculations were performed to generate data reported in this work. All described computations were performed in GAMESS-2014,[38] MOPAC-2016,[28,27,30] PM7-MD,[39] and GMSEARCH[40] quantum-chemistry codes. Where applicable, we used procedures, functions, and libraries from ASE and SciPy.org computational chemistry libraries.[41–43] Preparation of systems as well as the preparation of artwork were performed in VMD-1.9.1,[44] Gabedit-2.8,[45] and Avogadro-1.2.0 cross-platform software.[46]

**Results and Discussion**

The analysis of the potential energy surface even of small chemical systems provides invaluable information of their most thermodynamically stable local structure patterns. The configurations can be subsequently arranged according to their energetics to reveal the role of each structure in a continuous (condensed) phase of the corresponding substance. Whereas the long-range structure is normally sensitive to the size of the computer-simulated cluster, the local structure is trustworthy even if it is obtained for a single pair of molecules. Classical molecular dynamics simulations greatly benefited from this fact in the past decades as all empirical models (force fields) were derived based on extremely small clusters or, more frequently, gas-state molecules.

In this work, we used systems consisting of relatively small molecules and ions, to generate a vision of physical and chemical interactions in complexes of platinum and in their clusters with counterions where applicable. The reported results on covalent bonds and angles of the diethylsulfoxidopentachloroplatinate anion are directly comparable to the recently published X-ray experiment,[18] and in all cases the simulated structures look very reliable and consistent. Figure 1 represents the global-minimum configuration of the $[PtCl_5(DESO)]^-$ anion. The system was obtained by starting from the individual DESO molecule and the platinum (IV) chloride complex. Furthermore, additional simulations were used to investigate kinetic and thermodynamic stabilities of [PtCl$_5$(DESO)]$^-$ by starting from the state of diethylsulfoxidopentachloroplatinate.

The DESO molecule substitutes one of the chlorine atoms in [PtCl$_6$]$^{2-}$ and becomes the only neutral ligand in the platinum complex. Interestingly, the platinum-sulfur bond length is equal to the platinum-chlorine distance, whereas the Cl-Pt-S covalent angle is very close to 180 degrees and also to the Cl-Pt-Cl angles. Therefore, the symmetry of the platinum complex is insignificantly perturbed due to the ligand-exchange reaction. The exchanged chlorine atom is then energetically favorably bound to the proton then initially belonged to hexachloroplatinic acid in the experiments of Tkacheva and coworkers.[18]

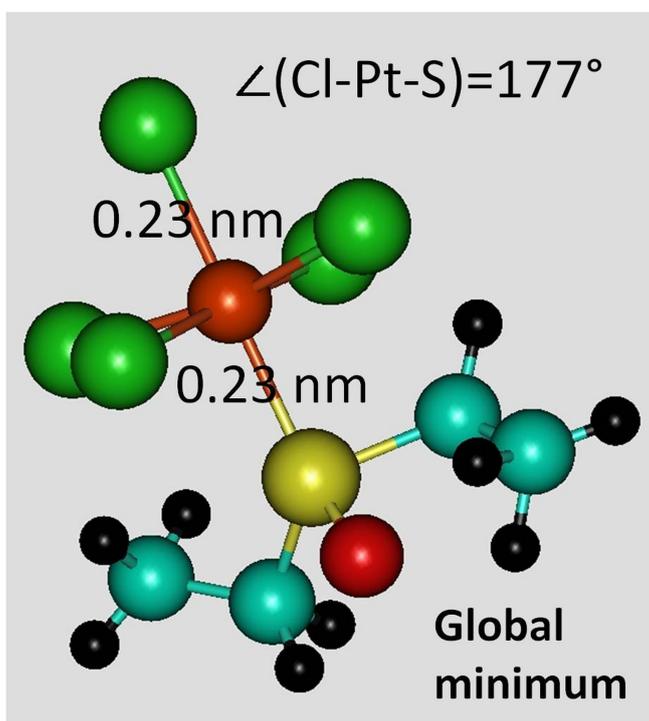

Figure 1. The global-minimum configuration of the [PtCl$_5$(DESO)]$^-$ anion obtained via a comprehensive scan of the potential energy surface. Oxygen atom is red, sulfur atom is yellow, carbon atoms are cyan, hydrogen atoms are black, chlorine atoms are green, and the platinum atom is orange.

It represents a natural fundamental interest whether more DESO-rich complexes of platinum are possible. Whereas the latter has not been reported experimentally yet, general chemical wisdom suggests that such complexes might be possible even though less thermodynamically stable. Indeed, the potential energy surface investigation of system 2 revealed the formation of the didiethylsulfoxidotetrachloroplatinate [PtCl$_4$DESO$_2$]$^0$ complex the global-minimum and local-minimum configurations of which are given in Figure 2.

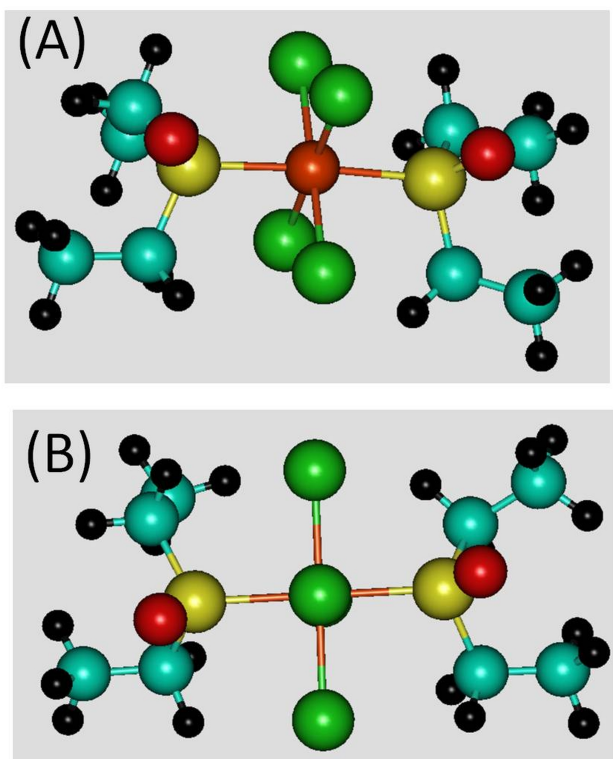

Figure 2. Stable molecular configurations of PtCl$_4$(DESO)$_2$ obtained upon a comprehensive search over the potential energy surface. (A) The global-minimum geometry. (B) The local-minimum conformation that is 4 kJ mol$^{-1}$ less thermodynamically stable than the global-minimum structure.

The DESO ligands of the [PtCl$_4$DESO$_2$] molecule are placed symmetrically forming a covalent angle S-Pt-S close to 180 degrees and comparable to the Cl-Pt-Cl and Cl-Pt-S angles. Such a molecule exhibits kinetic stability at 300 K and attains a few stationary-point configurations. The difference between configurations is 4 kJ mol$^{-1}$. It can be reliably associated with conformational flexibility of DESO, i.e. fluctuating non-covalent angles between the alkyl chains. Despite performing an extensive scan of conformations of [PtCl$_4$DESO$_2$], we were never able to observe any local-minima corresponding to the right angle S-Pt-S between the two DESO ligands. Thermodynamic and kinetic stabilities of [PtCl$_4$DESO$_2$] are expected to inspire experimental efforts to enlarge the universe of platinum-diethyl sulfoxide compounds.

Figure 3 exemplifies a search for even more DESO-rich complexes. The system with two DESO ligands and one DESO molecule was comprehensively investigated as described in Table 1. Unlike in the above cases, no instances of [PtCl$_3$(DESO)$_3$]$^+$ was observed. The global minimum of the system (Figure 3A) represents a weakly coordinated DESO molecule. The DESO molecule tends to approach one the methylene groups of one of the DESO ligands. The closest-approach distance between oxygen of the DESO molecule and hydrogen of the DESO ligand is 0.23 nm. The distance between platinum and sulfur (a potential donor-acceptor covalent bond) is 0.64 nm. The smallest distance between hydrogen of DESO and chlorine of the platinum complex is 0.29 nm. Local minima exhibit similar structure patterns, in no cases pointing to the possible emergence of a transition-state molecular structure or, at least, a decrease of the sulfur-platinum distance. Such univocal results reflect a hindrance along the anticipated reaction coordinate and call for the precise investigation of its origin and essence.

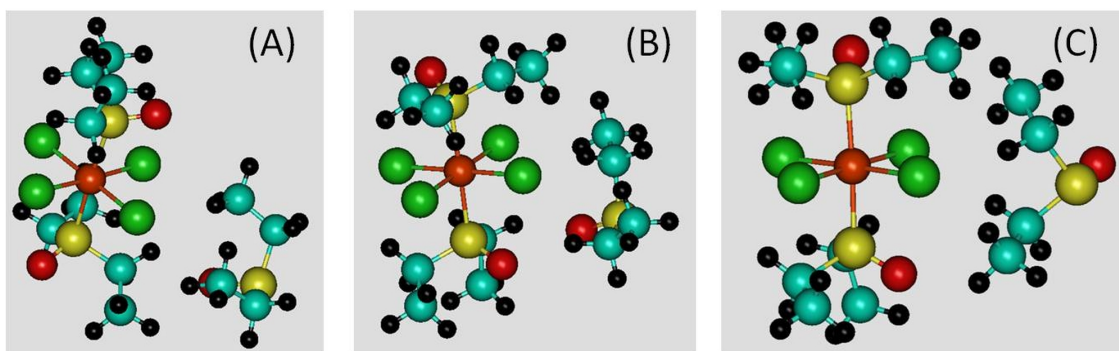

Figure 3. Selected stationary points of system 3. (A) The global minimum (point #4). (B) The local minimum (point #56), +10 kJ mol$^{-1}$. (C) The local minimum (point #84), +22 kJ mol$^{-1}$. The cation [PtCl$_3$(DESO)$_3$]$^+$ was searched for, but the attachment of DESO did not happen despite numerous iterations. The provided point numbers can be observed in Figure 6.

Since it was impossible to catch the [PtCl$_3$(DESO)$_3$]$^+$ cation upon the potential energy surface sampling, we attempted to create it forcibly. The distance between the sulfur atom of DESO and the platinum atom of the [PtCl$_4$(DESO)$_2$] non-charged complex was decreased gradually with a step of 0.01 nm. The procedure was started with low-energy stationary-point geometries of systems 3 and 4. At every step, the platinum and sulfur atoms were rigidly fixed at certain distance, while all other atoms in the system were allowed to relax. The steepest-descent geometry optimization was performed and the smallest potential energy was recorded for each Pt-S distance (Figure 4). When the Pt-S distance decreased from 0.59 nm to 0.56 nm, the potential energy of the system slightly decreased indicating a favorable direction of the steered motion. However, from 0.56 nm to 0.45 nm the potential energy rised resulting in a total penalty of +60 kJ mol$^{-1}$. This penalty is associated with a sterical hindrance created by the DESO ligands of [PtCl$_4$(DESO)$_2$]. In particular, collisions (distances of ~0.3 nm) were detected between the oxygen atom of the approaching DESO molecule and the DESO ligands. Further pulling of DESO towards the platinum atom led to the wave function optimization failure, i.e. the convergence of the iterative procedure with the required precision was impossible.

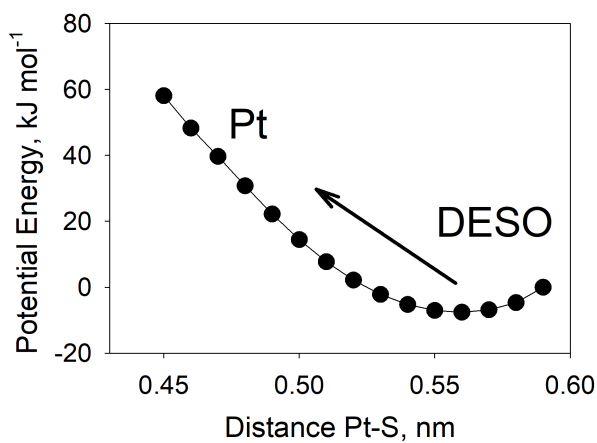

Figure 4. Relaxed potential energy scan along the reaction coordinate of attachment of the third DESO molecule to the platinum complex containing two DESO ligands.

To reiterate, formation of the $[PtCl_3(DESO)_3]^+$ cation is prohibited due to insufficient space around the platinum atom and the thermodynamically favorable conformations of the existing neutral DESO ligands. Unlike $[PtCl_4(DESO)_2]$, tridiethylsulfoxidotrichloroplatinate is not a viable compound. The same sterical restrictions were arguably observed in systems 5-7. The charge of the complex, the number of chlorine atoms, and the number of DESO ligands in no way influence the above described sterical restriction making $[PtCl_2(DESO)_4]^{2+}$, $[PtCl(DESO)_5]^{3+}$, and $[Pt(DESO)_6]^{4+}$ severely thermodynamically prohibited.

The melting point of tetraethylammonium diethylsulfoxidopentachloroplatinate amounts to 164°C.[18] Therefore, the novel compound is not a room-temperature ionic liquid by definition. Nevertheless, it remains an interesting product that combines organic and inorganic moieties. Analysis of the cation-anion binding in tetraethylammonium diethylsulfoxidopentachloroplatinate allows for understanding of its relatively high normal melting point and sheds light on the key interaction centers responsible for the physical-chemical properties. Figure 5 depicts the global-minimum and local-minimum configurations that are quite similar. In all cases, the cation prefers to bind to the chlorine atoms. This interaction is clearly electrostatically driven, whereas DESO represents a hydrophobic part to the anion. It may, therefore, foster solubility of tetraethylammonium diethylsulfoxidopentachloroplatinate in relatively non-polar solvents.

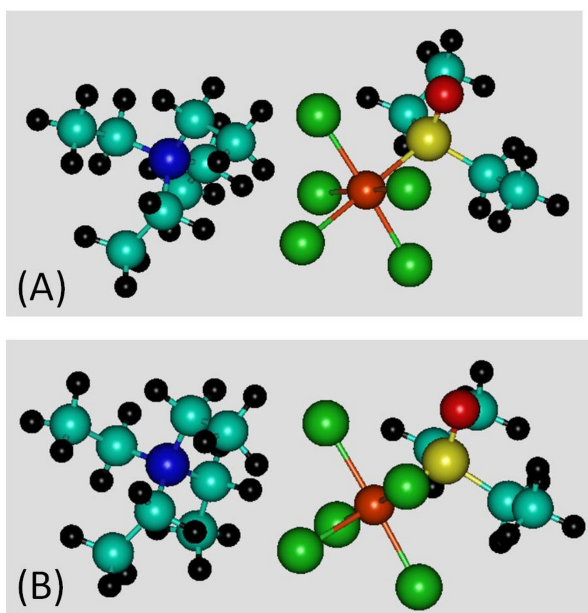

Figure 5. Selected stationary-point configurations of the tetraethylammonium diethylsulfoxidopentachloroplatinate ion pair. (A) Global-minimum geometry. (B) Local minimum that is by 0.8 kJ mol$^{-1}$ less thermodynamically favorable than the global minimum.

Conformational poorness of tetraethylammonium diethylsulfoxidopentachloroplatinate is convincingly supported by Figure 6. Indeed, the ion pair exhibits only six stationary-point configurations, whereas the energetic difference between them does not exceed 1 kJ mol$^{-1}$.

The existing conformational flexibility is evidently linked to the alternating angles between the alkyl chains of the cation and the neutral DESO ligand. Neither sulfur nor oxygen atoms play any role of the electron-rich interaction centers after the DESO molecule becomes a neutral ligand of the platinum complex. Conformational rigidity of the salt is in line with its above mentioned relatively high melting point of 164°C. Furthermore, it seems evident that substitution of the tetraethylammonium cation by a more coordinating species would result in a substantial extension of the solid phase temperature range. Unfortunately, our simulations are unable to directly predict the phase behavior of the salt, but it is possible to roughly rate the melting points of different salts based on the local cation-anion patterns. Notably, the conformational flexibility of system 4 (Figure 6) is much larger. The difference between the global-minimum configuration and the least thermodynamically competitive stable configuration amounts to 29 kJ mol$^{-1}$. This fact is rationalized by the presence of a non-bonded DESO molecule (Figure 3) and the absence of strong electrostatic attraction in the non-ionic system.

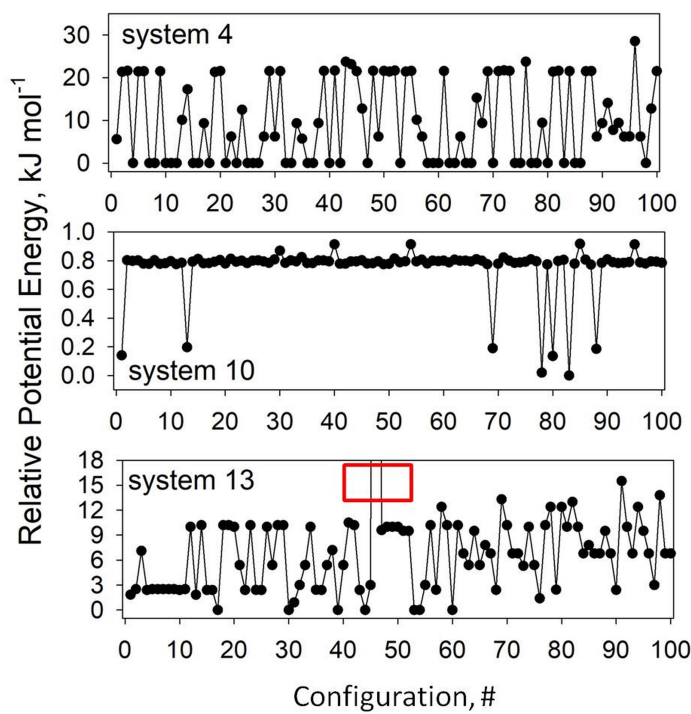

Figure 6. Potential energies corresponding to different stationary points on the potential energy surfaces in systems 4, 10, and 13. The red rectangle marks an occasionally identified transition state that corresponds to a complex dissociation (separation of ligands) process at high temperatures.

Thanks to the absence of strong electrostatic interactions among particles, [PtCl$_4$(DESO)$_2$] exhibits a substantial conformational flexibility. The difference between the most favorable minimum and the least favorable minimum is as large as 15 kJ mol$^{-1}$. Note a drastic difference with the ion-governed system 10.

Radial distribution functions for a few most interesting non-bonded atom pairs of the cation and the anion in system 11 are provided in Figure 7. The cation-anion strong binding is reflected by a distinct peak at the platinum-nitrogen RDF located between 0.47 and 0.53 nm. The distance agrees well with the sterical impossibility for the centers of the interacting ions to create a direct contact. The platinum-hydrogen (of (C$_2$H$_5$)$_4$N$^+$) RDF exhibits four peaks corresponding to the four ethyl chains. The first hydrogen maximum is located at the distance of 0.35 nm from the platinum atom. An alternative way to describe the cation-anion coordination is through the sulfur-nitrogen RDF. One notices that the corresponding S-N peak is located at a somewhat larger distance, 0.63 nm, than in the case of Pt-N RDF. This finding perfectly agrees with the conclusions derived upon the analysis of the selected stationary points (Figure 5). In particular, the sulfur atom belongs to a relatively non-polar moiety of the platinum complex. It is, therefore, reasonable that it is located farther away from the geometrical center of the cation.

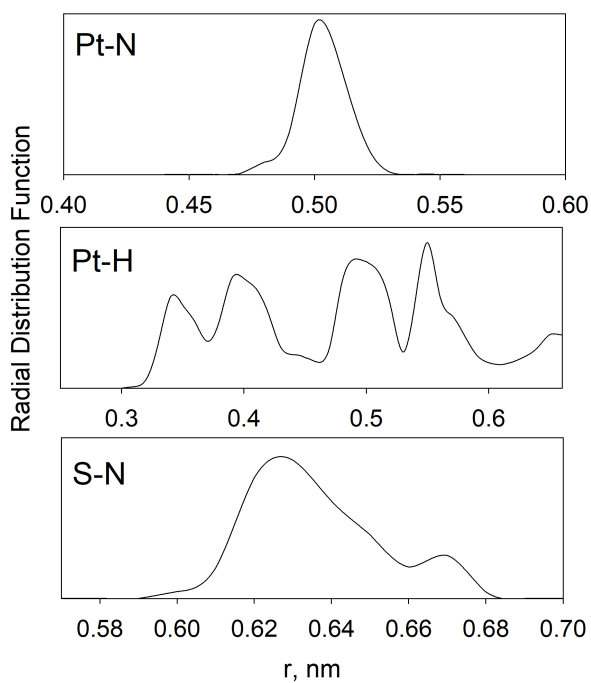

Figure 7. Radial distribution functions computed for the key interaction centers in system 11 to characterize cation-anion binding.

Since the recent paper supplies ready-to-use experimental data derived from the X-ray diffraction studies,[18] we made point-to-point comparisons for the most interesting structure parameters (Figure 8). Interestingly enough, the PM7 Hamiltonian provided a more accurate description of the new platinum-based complex structures than the hybrid density functional theory in the case of the platinum-sulfur donor-acceptor covalent bond. In other cases, the divergence of the results is negligible for this type of studies and can be safely ignored. An ability of the applied computational methods to produce high-quality interatomic distances enforces the reliance of all findings reported in the present work.

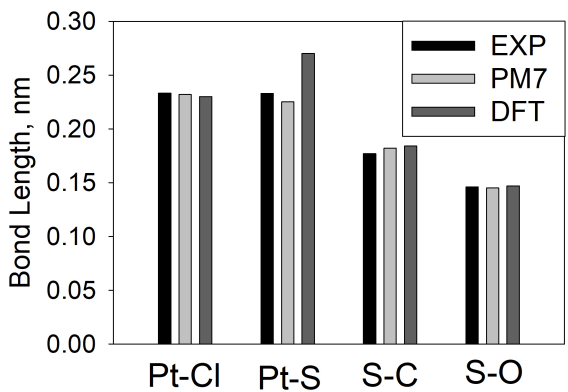

Figure 8. Comparison of the experimental covalent distances with the global-minimum bond lengths of the corresponding chemical bonds obtained in the present work.

The calculated infrared spectrum (Figure 9) successfully reproduces experimental numbers.[18] The most intensive bands correspond to the most polar covalent bonds in the system, S=O at 1160 cm$^{-1}$ and Pt-Cl at 310 cm$^{-1}$. The sulfur-platinum bond vibrates at 540 cm$^{-1}$ in the simulations, whereas the experimental peak is located at 530 cm$^{-1}$. The C-H bonds of the alkyl chains are expectedly fingerprinted beyond 3000 cm$^{-1}$, whereas the H-C-H bending occurs at ~1500 cm$^{-1}$. Overall, a decent reproduction of the major vibrational frequencies was achieved.

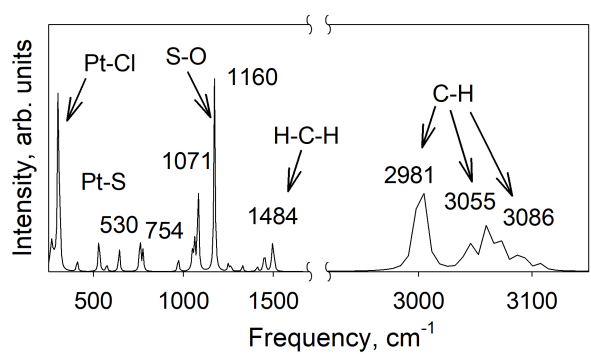

Figure 9. Infrared spectrum derived from the analysis of vibrational frequencies in the diethylsulfoxidopentachloroplatinate anion [PtCl$_5$(DESO)]$^-$. The numbers typed explicitly on the plot are taken from the experimental spectrum reported by Tkacheva and coworkers.[18]

**Conclusions**

This paper reveals new synthetic possibilities in coordination chemistry of platinum by providing thermodynamical argumentation of the stability of didiethylsulfoxidotetrachloroplatinate molecule. This structure clearly exhibits a kinetic stability at room temperature, whereas the corresponding stationary point on the potential energy surface was identified via the systematic search procedure. In turn, the tridiethylsulfoxidotrichloroplatinate cation and the tetradiethylsulfoxidodichloroplatinate cations are thermodynamically unstable due to sterical reasons. The performed simulations explicitly showed that these structures cannot be formed whatsoever because of collisions between the neutral DESO ligands and the approaching DESO molecule.

The cation-anion coordination in tetraethylammonium diethylsulfoxidopentachloroplatinate [(C$_2$H$_5$)$_4$N][PtCl$_5$(DESO] is rather strong as evidenced by the radial distribution functions for the respective central atoms. This finding explains an experimentally observed high melting temperature despite the presence of a weakly coordinating cation. Strong electrostatic attraction between the cation and the anion suppresses conformational flexibility in the studied salt and this is clearly seen upon the comparison with the simulated non-ionic systems. The specific results of this work will be interesting for researchers in the field of platinum-based coordination chemistry and dialkyl sulfoxides.

Stationary points on the potential energy surface correspond to the most chemically interesting molecular and ionic configurations that a system can attain. Systematic identification of the global and local minima provides a lot of useful chemical-physical information that can be normally extrapolated to the condensed phase of the corresponding substance. Conformational flexibility is an important descriptor that reflects the system's transport properties and allows to roughly hypothesize its phase diagram by comparing to the compounds of the same group. Analysis of the global and local minima makes it possible to directly rationalize many experimental findings and propose robust guidelines for novel synthetic endeavors.

**Acknowledgments**

**Conflict of interest**

The author hereby declares no financial interest.

**Author for correspondence**

All correspondence regarding the scientific content of this paper shall be directed through electronic mail to Prof. Dr. Vitaly V. Chaban (vvchaban@gmail.com).

4-17.